\documentclass[twocolumn,aps,prb,reprint,showpacs,amsmath,amssymb]{revtex4-1}
\usepackage{epsfig}
\usepackage{graphicx}
\usepackage{dcolumn}
\usepackage{bm}
\usepackage[colorlinks=true,dvipdfm]{hyperref}

\usepackage{bookmark}

\begin{document}
\title{Kibble-Zurek mechanism beyond adiabaticity: Finite-time scaling with critical initial slip}
\author{Yingyi Huang}
\author{Shuai Yin}
\email{sysuyinshuai@gmail.com}
\author{Qijun Hu}
\author{Fan Zhong}
\email{stszf@mail.sysu.edu.cn}
\affiliation{State Key Laboratory of Optoelectronic Materials and Technologies, School of Physics and Engineering, Sun Yat-sen University, Guangzhou 510275, People's Republic of China}
\date{\today}

\begin{abstract}
The Kibble-Zurek mechanism demands an initial adiabatic stage before an impulse stage to have a frozen correlation length that generates topological defects in a cooling phase transition. Here we study such a driven critical dynamics but with an initial condition that is near the critical point and that is far away from equilibrium. In this case, there is no initial adiabatic stage at all and thus adiabaticity is broken. However, we show that there again exists a finite length scale arising from the driving that divides the evolution into three stages. A relaxation--finite-time scaling--adiabatic scenario is then proposed in place of the adiabatic--impulse--adiabatic scenario of the original Kibble-Zurek mechanism. A unified scaling theory, which combines finite-time scaling with critical initial slip, is developed to describe the universal behavior and is confirmed with numerical simulations of a two-dimensional classical Ising model.
\end{abstract}
\pacs{64.60.De, 64.60.F-, 64.60.Ht, 05.70.Ln}

\maketitle

The Kibble-Zurek mechanism (KZM)~\cite{Kibble1,Kibble2,Zurek1,Zurek2,Zurek4} describes topological defect formation in driven critical dynamics in a variety of systems, ranging from classical~\cite{Laguna1,Laguna2,Laguna3,Laguna4,Laguna5,Laguna6,Laguna7,Laguna8,Laguna9,Laguna10,Laguna11,Laguna12,Laguna13,Laguna14,Laguna15,Vinas1,Vinas2,Vinas3,Vinas4,Vinas5,Vinas6,Vinas7,Liu} to quantum phase transitions~\cite{qkz1,qkz2,qkz3,qkz4,qkz5,qkz6,qkz7,qkz8,qkz9,qkz10,revqkz1,revqkz2}. Kibble first proposed it in cosmology~\cite{Kibble1,Kibble2} by identifying a frozen correlation length $\hat{\xi}$ that renders spatially distant regions causally independent during the cooling of the universe from the big bang. Then, Zurek brought this proposal to condensed matter physics and offered a method to compute the density of defects formed~\cite{Zurek1,Zurek2}. As a system cannot always follow adiabatically the cooling of a finite rate $R$ due to critical slowing down near the critical point, its evolution from a temperature $T_0$, sufficiently higher than the critical temperature $T_c$, can be divided into three sequential stages, an initial adiabatic stage, an impulse stage, and a final adiabatic stage below $T_c$. In the initial adiabatic regime, the correlation length $\xi$ and the correlation time $\zeta_s$ grow as $|\varepsilon|^{-\nu}$ and $\xi^z$, respectively, as the distance to the critical point, $\varepsilon\equiv T-T_c$, is reduced, where $\nu$ and $z$ are the correlation-length and the dynamic critical exponents, respectively~\cite{Cardy}. The boundaries between the stages are then determined by the frozen instant $\hat{t}$ at which the time interval before the transition, $t_c-t= \varepsilon/R$, equals $\zeta_s$~\cite{Zurek1,Zurek2}, where $t_c=\varepsilon_0/R$ is the time at $\varepsilon=0$. This leads to $t_c-\hat{t}\sim R^{- z/r_T}$~\cite{Zurek1,Zurek2}, where $r_T=z+1/\nu$ is a rate exponent~\cite{Zhongjp6}. Upon assuming evolutionless in the middle impulse stage, $\hat{t}$ then determines $\hat{\xi}\sim R^{-1/r_T}$ and thus the defect density $n\sim R^{d/r_T}$, the KZ scaling~\cite{Zurek1,Zurek2}.

Crucial in the derivation is the existence of the initial adiabatic stage that gives rise to $\hat{\xi}$. It results from the large $\varepsilon_0=T_0-T_c$ and thus small $\zeta_s$. By contrast, whether the initial state is equilibrium or not is irrelevant as the system can quickly equilibrate once $\zeta_s$ is small. This has been confirmed by a lot of experiments and numerical simulations~\cite{Haller,DeGrandi,Dengss,Vinas3,Vinas4,Vinas5,Zurek3,Zurek5}. On the other hand, when $\varepsilon_0=0$ and the initial state is the equilibrium state there, it has been shown by an adiabatic perturbation method that the scaling of topological defects is consistent with the KZ scaling~\cite{Polkov1,Polkov2,revqkz1}. Yet, it is difficult to obtain the equilibrium state near the critical point due to critical slowing down.

However, if $\varepsilon_0$ is small and the initial state is not the equilibrium state at $\varepsilon_0$, the equilibration of the system has to take a long time as the relaxation time $\zeta_s$ is now macroscopically large. In this case, there is no initial adiabatic stage at all and thus adiabaticity is broken. Questions then arise as to whether there is still a $\hat{\xi}$ that generates topological defects, or, whether the KZM is still valid or not. Does universal behavior exist in this driving critical system with a nonequilibrium initial condition? If the answers are yes, then how is $\hat{\xi}$ determined and how does one describe the universal behavior, as the adiabatic--impulse--adiabatic scenario of the KZM cannot apparently be applied to this case?

Relaxation of a nonequilibrium initial state near $\varepsilon_0$ is not a strange situation~\cite{Hohenberg,Tauber,Calabrese}. A well-known case is the critical initial slip~\cite{Janssen1,Janssen2}, which was found in classical~\cite{Janssen1,Janssen2} and recently in quantum critical phenomena in imaginary time~\cite{Yin}. When a system is quenched rapidly from a high temperature disordered state to near its critical point and relaxes, it has been found that the order parameter $M$ grows as $M\sim M_0t^\theta$ right after a microscopic timescale, where $\theta$ is an independent initial-slip exponent and $M_0$ is a small initial order parameter, which may be generated by an external field~\cite{Janssen1,Janssen2,Huse89}. As the initial state is derived from the disordered phase, it possesses only short-ranged correlations. However, finite-ranged correlations are irrelevant in the renormalization-group (RG) sense~\cite{Janssen1,Janssen2}. So, the initial state may be an equilibrium state of a Hamiltonian different from the system's.

A system that is driven by an external field including the temperature with a constant time rate $R$ through its critical point is well described by the theory of finite-time scaling (FTS)~\cite{Zhong1,Zhong2}. It is a temporal counterpart of the well-know finite-size scaling~\cite{Cardy} and is derived from the RG theory~\cite{Zhong1,Zhong2}. FTS shows that there is a finite timescale $\zeta_d\sim R^{-z/r_T}$ induced by an external driving; and when $\zeta_d$ is shorter than $\zeta_s$, it dominates the evolution in an FTS regime. This indicates that the impulse regime of the KZM is just an FTS regime. Indeed, $\zeta_d$ is just $\hat{t}$ because at this instant $\zeta_d=\zeta_s$ and $\hat{\xi}$ is just the length scale corresponding to $\zeta_d$. Moreover, the scaling behavior in the evolutionless impulse regime and the KZ scaling are well described by FTS~\cite{qkz10,Yin3,Huang,Yin2}. In FTS, however, the initial state is, similar to the KZM, far away from the critical point and has thus no effects.

Here, in order to describe the scaling behavior of a driven critical system with a nonequilibrium initial condition, we combine FTS with the critical initial slip. We shall show that there again exists in this case the finite timescale $\zeta_d$ and thus $\hat{\xi}$. As a result, the KZM for topological defect formation is still valid though  adiabaticity is broken. However, its adiabatic--impulse--adiabatic scenario is now changed to a relaxation--FTS--adiabatic scenario, in which a nonequilibrium nonadiabatic relaxation stage replaces the original initial adiabatic stage. In this relaxation stage, the growing correlation time $\zeta_i$, which is different from $\zeta_s\sim|\varepsilon|^{-\nu z}$, dominates the evolution and $\zeta_d$ is subsidiary. Once $\zeta_i$ gets longer than $\zeta_d$, the latter takes over and the system enters the FTS stage. This is the impulse stage of the KZM. However, in the KZ sense, both the relaxation and the FTS stages are impulse as both are nonadiabatic albeit due to different reasons, viz., the former arises from the initial conditions whereas the latter from the driving. When the system is driven to so far away from $T_c$ that $\zeta_s$ becomes shorter than $\zeta_d$, it crossovers into the adiabatic stage.

\begin{figure}
  \centerline{\epsfig{file=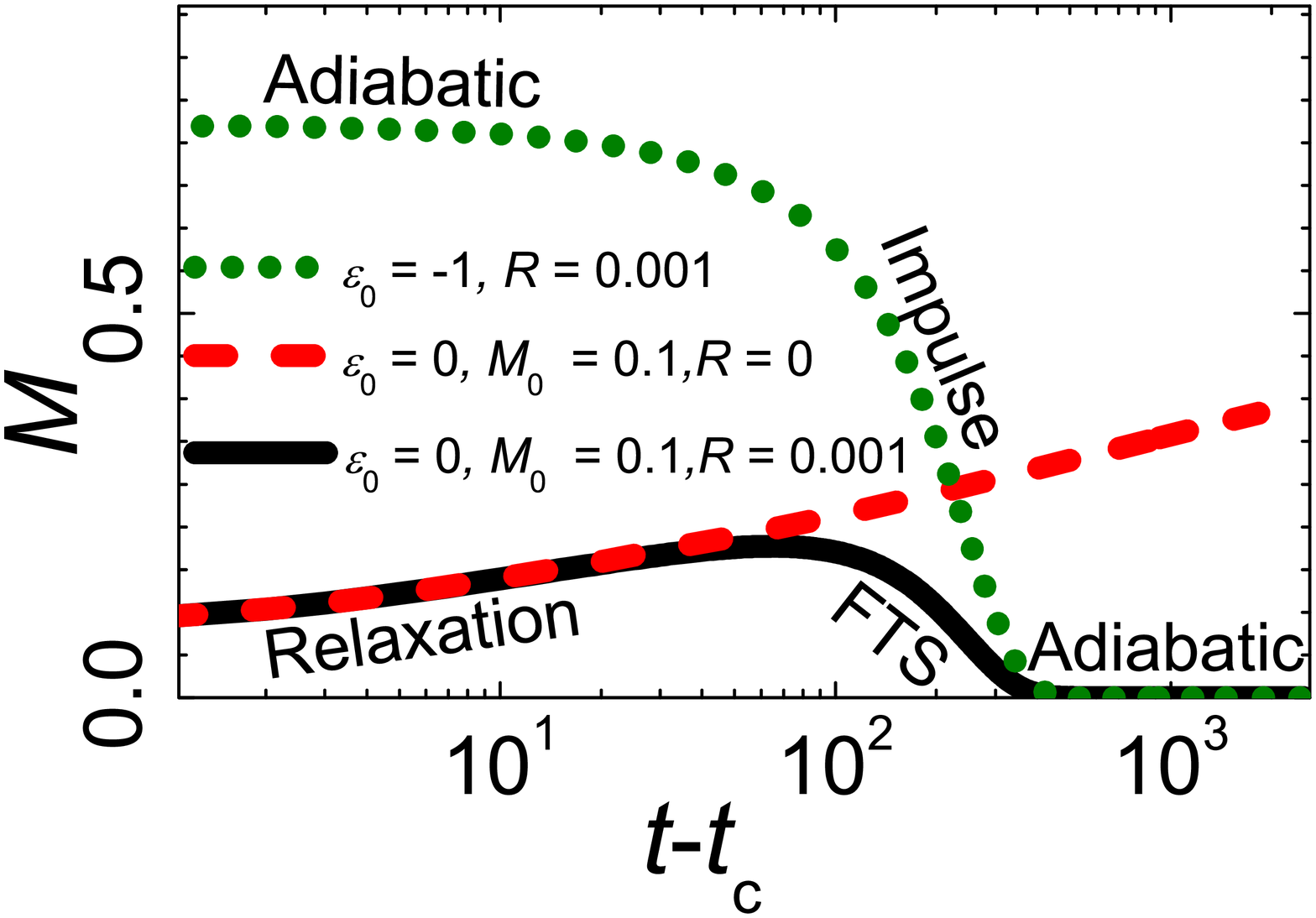,width=0.7\columnwidth}}
  \caption{\label{ftsqcr}(Color online) Evolution of the order parameter $M$ for the two-dimensional classical Ising model. The initial distance to the critical point $\varepsilon_0$ and the initial order parameter $M_0$ are indicated except the one with $\varepsilon_0=-1$, where any given $M_0$ will always decay rapidly to the equilibrium values. Heating instead of cooling is performed to facilitate presentation. Different stages are marked.}
\end{figure}

As an appreciation of the results, we plot in Fig.~\ref{ftsqcr} the evolution of $M$ for three different sets of the initial conditions. One sees that the evolution starting with a large $|\varepsilon_0|$ and at $T_c$ show qualitatively distinct behavior. Driving ($R\neq0$) makes no appreciable difference at short times when the system starts with an uncorrelated nonequilibrium initial state near to its $T_c$. In this stage, since the correlation length grows as $\xi_i\sim t^{1/z}$ and $\zeta_i\sim\xi_i^z$, $\zeta_i\sim t$~\cite{Janssen1,Janssen2}. As $\zeta_i$ is shorter than $\zeta_d$ in short times, it dominates the dynamics and the stage is thus relaxational similar to the critical initial slip, while the external driving in only a perturbation. Note that this relaxation stage has nothing to do with the free relaxation regime~\cite{Vinas3,Vinas6,Vinas7} that follows the final adiabatic stage and that has no driving at all.

In the following, we shall first present the scaling theory and obtain different scaling behaviors in different stages and their crossovers. These are then confirmed by simulations on a two-dimensional ($2$D) classical Ising model. As this is a generic model for critical phenomena, we expect the results to be applicable to other models and even to quantum critical behavior as well. We shall not study the topological defects as their counting is not easy and detecting scaling behavior of other observables has been advocated~\cite{Zurek3,Chandran}.

Near the critical point, the scaling behaviors of macroscopic quantities can be readily described by a scale transformation. Our scaling theory is based on
\begin{eqnarray}
\begin{aligned}
&M(t,R,M_0,\varepsilon_0,\varepsilon)\\&=b^{-\beta/\nu}M(tb^{-z},Rb^{r_T},U(M_0,b),\varepsilon_0 b^{1/\nu},\varepsilon b^{1/\nu}). \label{scalingt}
\end{aligned}
\end{eqnarray}
for a rescaling of a factor $b$, where $|\varepsilon_0|\ll1$, $\beta$ is the critical exponent for $M$, and $U(M_0,b)$ is the universal characteristic function describing the rescaled initial magnetization~\cite{Zheng,Yin4}. In Eq.~(\ref{scalingt}), we have purposely written $t$, $R$, $\varepsilon_0$, and $\varepsilon$ out though they are not independent as $\varepsilon=\varepsilon_0+Rt$. For a small $M_0$, $U(M_0,b)=M_0 b^{x_0}$ with $x_0$ being the scaling dimension of $M_0$ \cite{Janssen1,Janssen2}. $U(M_0,b)$ has one fixed point $U(0,b)=0$ for arbitrary $b\geq 1$. For a hard-spin system, in which $M$ is bounded, the saturated $M_0$ is another fixed point, since the rescaled $M_0$ is invariant under coarse graining~\cite{Zheng}. Note that the two additional scaling variables, $\varepsilon_0$ and $M_0$, are present only for small $|\varepsilon_0|$; for large $|\varepsilon_0|$, they are absent as they are then irrelevant. Equation~(\ref{scalingt}) with a small $M_0$ can be justified by an RG theory which combines the critical initial slip~\cite{Janssen1,Janssen2} and the FTS theory \cite{Zhong1,Zhong2,Zhongjp6}.

The scaling forms of different stages can now be obtained from Eq.~(\ref{scalingt}) by comparing the relevant time scales. In the first stage, in which $t$ is small, $\zeta_i$ is small and growing. Accordingly, relaxation dominates. By setting $tb^{-z}=1$, we arrive at the scaling form
\begin{equation}
M(t,R,M_0,\varepsilon_0)=t^{-\beta/\nu z}f_1(Rt^{r_T/z},U(M_0,t^{1/z}),\varepsilon_0 t^{1/\nu z}), \label{scalingf1}
\end{equation}
where $f_1$ is a scaling function. It is valid when all scaled variables are small. In particular, $Rt^{r_T/z}\ll1$, or $t\ll R^{-z/r_T}$, i.e., $\zeta_i\ll\zeta_d$ as ought to be. Detailed scaling behavior can be obtained from Eq.~(\ref{scalingf1}) as follows.

For $\varepsilon_0=0$ and a small $M_0$, $U(M_0,t^{1/z})=M_0 t^{x_0/z}$. One can expand $f_1$ in $Rt^{r_T/z}$ and $M_0 t^{x_0/z}$ to the second order and obtains
\begin{equation}
M\simeq M_0t^\theta f_1^{'}(0,0,0)+t^{\theta+r_T/z}R M_0 f_1^{''}(0,0,0). \label{shortt}
\end{equation}
where $\theta=(x_0-\beta/\nu)/z$ and a prime stands for a partial derivative. In Eq.~(\ref{shortt}), the first term describes the usual critical initial slip~\cite{Janssen1,Janssen2}. The second term of Eq.~(\ref{shortt}) displays the driving-induced deviation from the critical initial slip. It is a mixed term between $M_0$ and $R$ and arises from the fact that, if $M_0=0$, $M$ remains zero as $\varepsilon$ does not break the symmetry. The external driving, which dominates near the critical point in the ordinary KZM, here acts only as a perturbation.

For $\varepsilon_0=0$ and the saturated $M_0$, $U(M_0,t^{1/z})=M_0$. In the initial stage, $M$ now decays according to
\begin{equation}
M\simeq t^{-\beta/\nu z}f_1(0,M_0,0)+Rt^{(\nu r_T-\beta)/\nu z}f_1^{'}(0,M_0,0), \label{shortt1}
\end{equation}
where the first term is the nonequilibrium relaxation~\cite{Ito} and the second term arises again from the perturbation of the driving. Note that for $R>0$, $f_1^{'}(0,M_0,0)<0$ because $M$ must decrease as the temperature increases.

Crossover to the FTS stage occurs at $R\hat{t}_i^{r_T/z}\sim 1$, or $\hat{t}_i\sim R^{-z/r_T}\sim\zeta_d$. This is not $\hat{t}$ of the KZM as it is the crossover from the relaxation stage, which is also nonadiabatic. However, the asymptotically identical forms show that $\zeta_d$ of the FTS regime does not depend on the initial conditions. The scaling form of the FTS stage can be obtained from Eq.~(\ref{scalingt}) as
\begin{equation}
M=R^{\beta/\nu r_T}f_2(\varepsilon_0 R^{-1/\nu r_T},U(M_0,R^{-1/r_T}),\varepsilon R^{-1/\nu r_T}) \label{scalingf2}
\end{equation}
with another scaling function $f_2$. For $\varepsilon_0=0$ and the saturated $M_0$, Eq.~(\ref{scalingf2}) is quite similar to the usual FTS form~\cite{Zhong1,Zhong2,qkz10}. However, the scaling functions are different, because they characterize different evolutions from distinct initial conditions as can be seen from Fig.~\ref{ftsqcr}.

When $\varepsilon R^{-1/\nu r_T}\gg1$, or $\zeta_s\ll\zeta_d$, the system enters the adiabatic stage with a scaling form
\begin{equation}
M(R,M_0,\varepsilon_0,\varepsilon)=\varepsilon^{\beta}f_3(\varepsilon_0\varepsilon^{-1},U(M_0,\varepsilon^{-\nu}),R\varepsilon^{-\nu r_T}), \label{scalingf3}
\end{equation}
where $f_3$ is a scaling function. This crossover is similar to the usual impulse--adiabatic crossover in KZM as can be seen in Fig.~\ref{ftsqcr}. Indeed, the time when the curve of the usual KZM tends to zero is close to the corresponding time of the curve starting with an nonequilibrium state. This indicates again that the timescale in the FTS stage is consistent with the timescale in the impulse region.

The scaling theory is applicable to other situations in which other variables than $T$ are changed starting with a nonequilibrium initial state near the critical point. For example, consider changing the symmetry-breaking field $h$ as $h=h_0+R_h t$ with a small $h_0$ and a constant $R_h$. We set $\varepsilon=0$ to reduced competing scales. In this case, there exist also three stages in the driving process. In the relaxation stage, the scaling form for small $M_0$ is
\begin{equation}
M=t^{-\beta/\nu z}f_{1h}(R_ht^{r_h/z},M_0 t^{x_0/z},h_0 t^{\beta \delta/\nu z}), \label{scalinghf1}
\end{equation}
with $r_h=z+\beta\delta/\nu$~\cite{Zhong1,Zhong2}, while in the FTS stage in which $\zeta_i\gg\zeta_d\sim R^{-z/r_h}$, the scaling form changes to
\begin{equation}
M=R_h^{\beta/\nu r_h}f_{2h}(M_0R_h^{-x_0/r_h},h_0 R_h^{-\beta \delta/\nu r_h},h R_h^{-\beta \delta/\nu r_h}), \label{scalinghf2}
\end{equation}
where $f_{1h}$ and $f_{2h}$ are scaling functions. Finally comes the $h$ dominated adiabatic stage. Again, $f_{2h}$ for $h_0=0$ and a saturated $M_0$ is different from the usual one with adiabatic initial conditions. We note that this case has been considered in Ref.~\onlinecite{Zhongjp6}, where a method to determine the critical exponents was proposed. However, the relaxation has not been discussed there.

To confirm the scaling theory, we take the $2$D classical Ising model as an example. Its Hamiltonian is
\begin{equation}
H=-\sum_{<i,j>}S_i S_j-h\sum_{i}S_i,
\label{isingh}
\end{equation}
where $S_i=\pm1$ and the first sum is over all nearest neighbors and the second over all spins. Note that unless changing the symmetry-breaking external field $h$, we set $h=0$ for simplicity. The critical point of (\ref{isingh}) is $T_c=2/\log(\sqrt{2}+1)$ \cite{Cardy} and the critical exponents are $\beta=1/8$, $\nu=1$, $\delta=15$~\cite{Cardy}, $z=2.1667$~\cite{Hohenberg}, and $\theta=0.191$~\cite{Li1,ZhengB,Li2}. They will be taken as inputs to verify the scaling forms.
The single-spin Metropolis algorithm~\cite{Binder} is used. The lattice size is $5000$, which has been checked to produce negligible size effects. Periodic boundary conditions are applied throughout. We calculated averages over between $2000$ and $3000$ samples, which guarantee that the relative uncertainty is  smaller than $1\%$. The initial configuration is a uniformly-distributed random assignment of $S_i=\pm1$ with an average equal to $M_0$.

\begin{figure}[t]
  \centerline{\epsfig{file=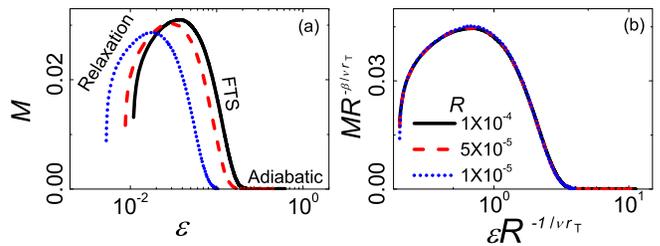,width=1.0\columnwidth}}
  \caption{\label{res2} (Color online) Three stages of the evolution of $M$ under increasing $\varepsilon$ with fixed $\varepsilon_0R^{-1/\nu r_T}=0.2$ and $M_0R^{-x_0/r_T}=0.06$ for three $R$ indicated. The curves before and after rescaled are shown in (a) and (b) respectively.  Semi-logarithmic scales are used.}
 \end{figure}
Firstly, we classify the different stages of the evolution and examining the scaling form~(\ref{scalingf2}) for small $M_0$. Figure~\ref{res2} shows the dependence of $M$ on $\varepsilon$ for several $M_0$ and $\varepsilon_0>0$. When $\varepsilon$ is small, $M$ increases with $\varepsilon$ and thus $t$ at short times. This is similar to the critical initial slip in the pure relaxation and is thus the relaxation stage. When $\varepsilon$ gets larger, $M$ decreases as $\varepsilon$ increases. Yet, $M$ increases with $R$ and hysteresis occurs. This is the generic behavior of FTS stage~\cite{Zhong1,Zhong2}. Then follows the adiabatic stage, in which $M$ is zero, independent of $R$ and the initial condition. Because $M_0$ and $\varepsilon_0$ have been chosen in such a way that $M_0R^{-x_0/ r_T}$ and $\varepsilon_0R^{-1/\nu r_T}$ are fixed, the curves collapse onto each other after rescaling according to Eq.~(\ref{scalingf2}), confirming that $M_0$ and $\varepsilon_0$ are indispensable scaling variables.

\begin{figure}[t]
  \centerline{\epsfig{file=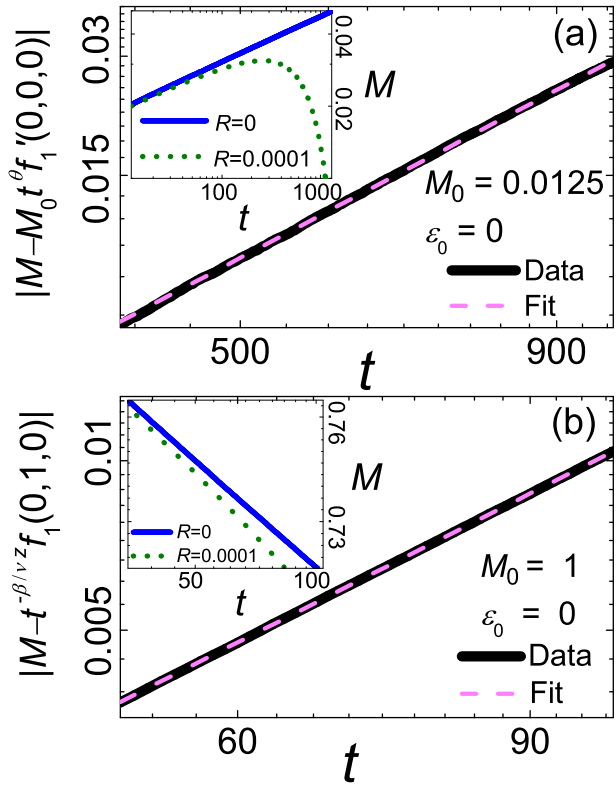,width=1.0\columnwidth}}
  \caption{\label{res3} (Color online) Difference of $M$ between $R\neq0$ and $R=0$ with (a) a small $M_0$ and (b) the saturated $M_0$ for $\varepsilon_0=0$. Double-logarithm scales are used. The insets plot the original curves, from which one sees that the initial slip emerges after ten Monte Carlo steps per spin or so.}
\end{figure}
Secondly, we study the effects of the external driving according to Eqs.~(\ref{shortt}) and (\ref{shortt1}). In Fig.~\ref{res3}(a), the difference between the driving relaxation and pure relaxation satisfies a power-law relation, $|M-M_0t^\theta f_1^{'}(0,0,0)|\propto t^{\theta+r_T/z}$, according to Eq.~(\ref{shortt}). The fitted slope is $\theta+r_T/z=1.696(2)$, which agrees with the theoretical value of $\theta+r_T/z=1.652$. For the case of the saturated $M_0$, which is $M_0=1$ for Ising model, as shown in Fig.~\ref{res3}(b), $|M-t^{-\beta/\nu z}f_1(0,1,0)|$ changes with $t$ with an exponent $1.499(4)$, which is close to $(\nu r_T-\beta)/\nu z=1.403$, consistent with Eq.~(\ref{shortt1}). The deviations arise from the contributions of higher order terms in the expansions.

Thirdly, we further verify the scaling theory by examining the scale transformation~(\ref{scalingt}) for large $M_0$.  In this case, the rescaled initial order parameter $M_0^{'}=U(M_0,b)$ is not a simple power-law~\cite{Zheng,Yin4}. So, for a given $b$ and $\varepsilon_0=0$ for instance, we first estimate $M_0^{'}$ from the pure relaxation by select an $M_0^{'}$ starting with which the evolution of $M$ matches, \begin{figure}[h]
  \centerline{\epsfig{file=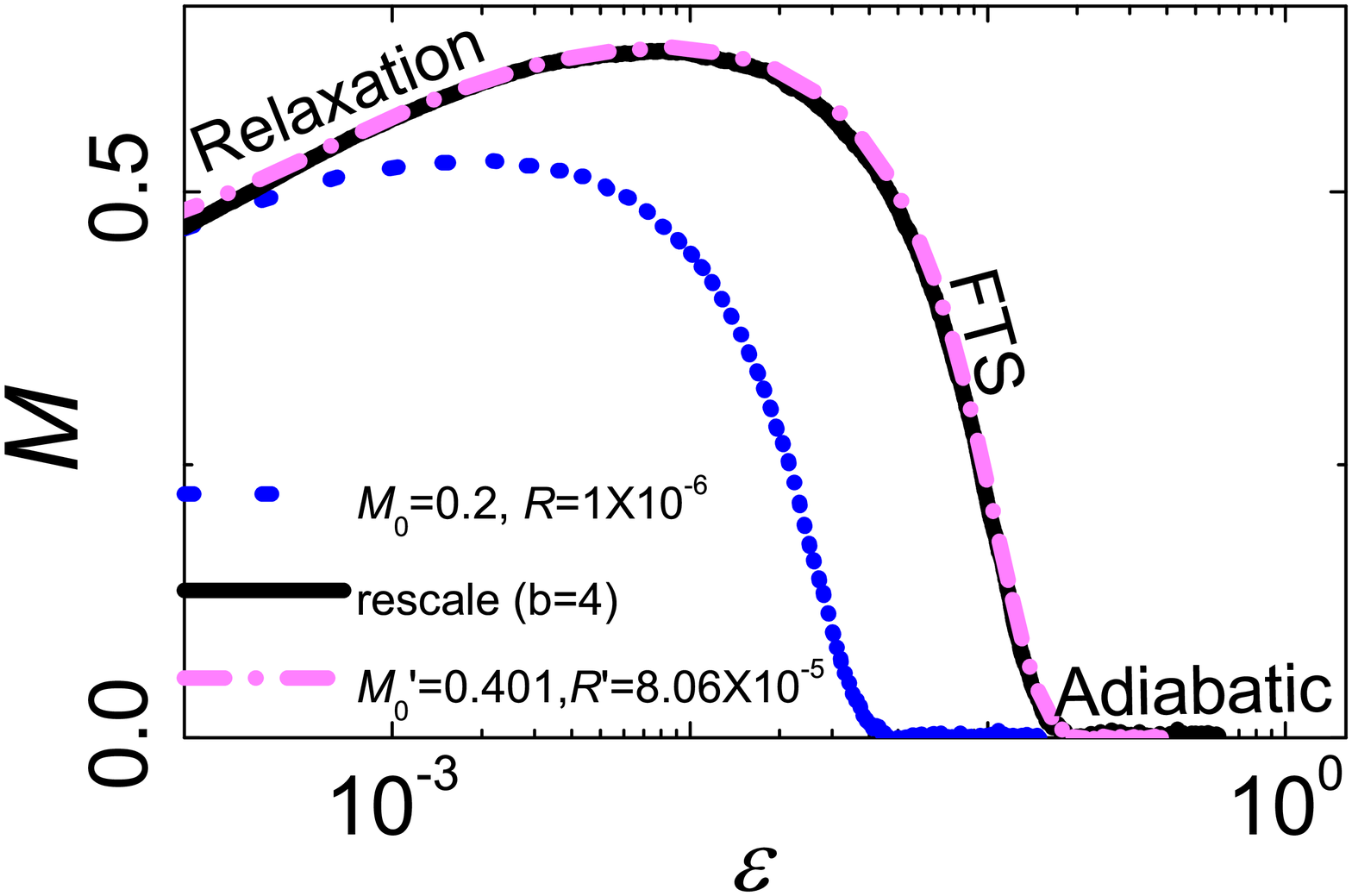,width=0.7\columnwidth}}
  \caption{\label{res6} (Color online) The evolution of $M$ under changing $\varepsilon$ with $\varepsilon_0=0$ and $M_0=0.2$. It matches the rescaled curves with the estimated value of $M_0^{'}$ and $R^{'}=Rb^{r_T}$ for $b=4$. Semi-logarithmic scales are used.}
\end{figure}
after its $M$ and $t$ being rescaled by $b^{-\beta/\nu}$ and $b^{-z}$, respectively, that starting with $M_0$~\cite{Zheng,Yin4}. With this $M_0^{'}$, the evolution of $M$ when $\varepsilon$ is changing again matches well that starting with $M_0$ upon proper rescaling, including $R^{'}=Rb^{r_T}$, as is illustrated in Fig.~\ref{res6}. Also manifest in the figure is the three stages similar to the case of small $M_0$. These show that the effects of driving and of the initial conditions are independent and thus confirm Eq.~(\ref{scalingt}).

\begin{figure}[t]
  \centerline{\epsfig{file=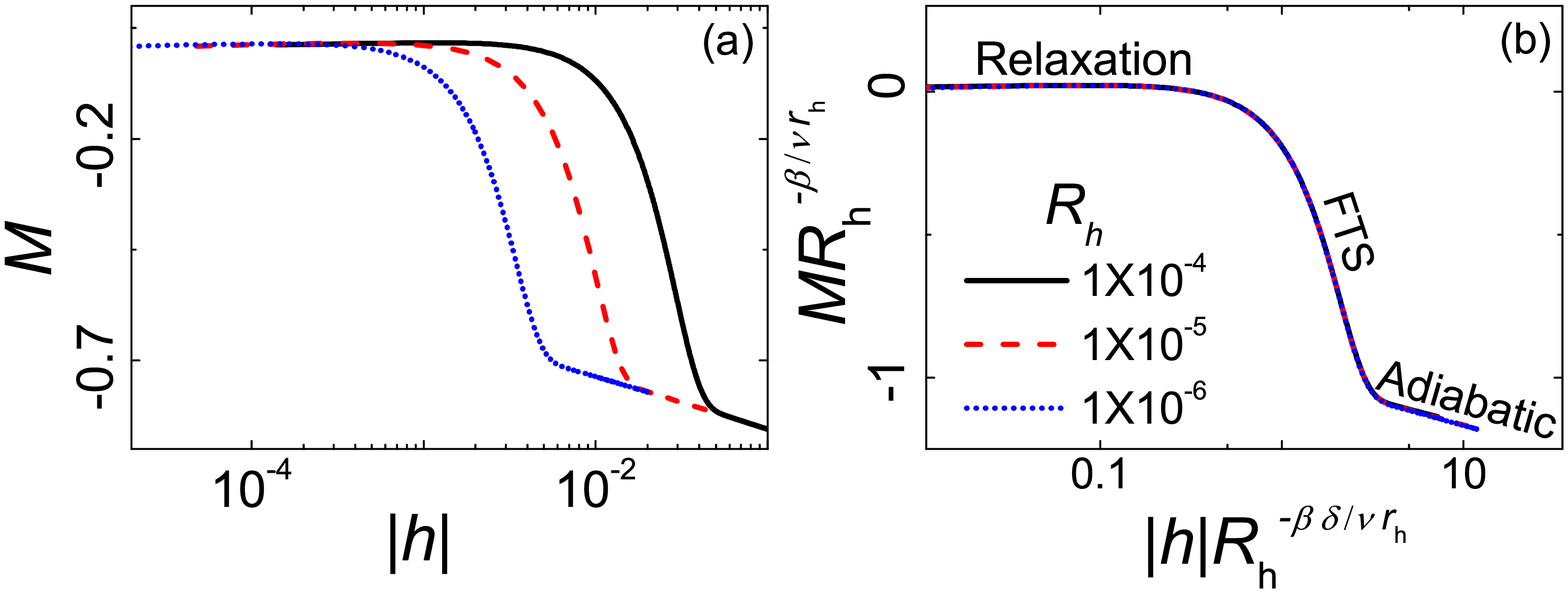,width=1.0\columnwidth}}
  \caption{\label{res5} (Color online) (a) Three stages of the evolution of $M$ under changing $h$ with fixed $|h_0|R_h^{-\beta\delta/\nu r_h}=0.01$ and $M_0R_h^{-x_0/ r_h}=0.04$ for three $R_h$ indicated. (b) The rescaled curves. Semi-logarithmic scales are used.}
\end{figure}
Fourthly, we consider the situation of changing the symmetry-breaking field $h$. Figure~\ref{res5} shows the results of changing $h$ as $h=h_0-R_ht$ with some small and negative $h_0$. The three stages also show manifestly similar to those in Fig.~\ref{res2}. The rescaled curves with different $R_h$ and $M_0$ collapse well onto each other for fixed $M_0R_h^{\-x_0/r_h}$ and $|h_0|R_h^{-\beta\delta/\nu r_h}$. This confirms both that the scaling form must include $h_0$ and $R_h$ as scaling variables as Eq.~(\ref{scalinghf2}) indicates and that the relaxation--FTS--adiabatic scenario is generally applicable in the driving dynamics with nonequilibrium initial states near the critical point.

In summary, we have systematically studied the driving dynamics starting with a nonequilibrium initial state near the critical point. This initial condition breaks the adiabaticity and thus changes the adiabatic--impulse--adiabatic scenario of the KZM into the relaxation--FTS--adiabatic scenario by suppressing the initial adiabatic stage. A scaling theory that combines FTS with critical initial slip has been developed and account well for the universal scaling behavior in this nonequilibrium nonadiabatic case. Numerical simulations on the 2D Ising model have confirmed that the theory applies well both to varying temperature and to varying the symmetry-breaking external field. Our theory might provide a way of nonadiabatic quantum computations as opposed to the adiabatic ones~\cite{Farhi}, as one may now quench nonadiabatically from the ground state of an initial Hamiltonian to the targeted one even at the critical point of the latter.


This project was supported by NNSFC (10625420).


\begin{thebibliography}{99}
\bibitem{Kibble1} T. W. B. Kibble, J. Phys. A: Math. Gen. {\bf 9}, 1387 (1976).
\bibitem{Kibble2} T. W. B. Kibble, Phys. Today {\bf 60} (9), 47 (2007).
\bibitem{Zurek1} W. H. Zurek, Nature {\bf 317}, 505 (1985).
\bibitem{Zurek2} W. H. Zurek, Phys. Rep. {\bf 276}, 177 (1996).
\bibitem{Zurek4} A. del Campo and W. H. Zurek, Int. J. Mod. Phys. A {\bf 29}, 1430018 (2014).
\bibitem{Laguna1} P. Laguna and W. H. Zurek, Phys. Rev. Lett. {\bf 78}, 2519 (1997).
\bibitem{Laguna2} A. Yates and W. H. Zurek, Phys. Rev. Lett. {\bf 80}, 5477 (1998).
\bibitem{Laguna3} N. D. Antunes, L. M. A. Bettencourt, and W. H. Zurek, Phys. Rev. Lett. {\bf 82}, 2824 (1999).
\bibitem{Laguna4}  M. Hindmarsh and A. Rajantie, Phys. Rev. Lett. {\bf 85}, 4660 (2000).
\bibitem{Laguna5}  I. L. Chuang, R. Durrer, N. Turok, and B. Yurke, Science {\bf 251}, 1336 (1991).
\bibitem{Laguna6}  M. J. Bowick, L. Chandar, E. A. Schiff, and A. M. Srivastava, Science {\bf 263}, 943 (1994).
\bibitem{Laguna7}  V. M. H. Ruutu, V. B. Eltsov, A. J. Gill, T. W. B. Kibble, M. Krusius, Yu G. Makhlin, B. Placais, G. E. Volovik, and W. Xu, Nature {\bf 382}, 334 (1996).
\bibitem{Laguna8} C. B\"{a}uerle, Yu M. Bunkov, S. N. Fisher, H. Godfrin, and G. R. Pickett, Nature {\bf 382}, 332 (1996).
\bibitem{Laguna9}  R. Monaco, J. Mygind, and R. J. Rivers, Phys. Rev. Lett. {\bf 89}, 080603 (2002).
\bibitem{Laguna10}  R. Carmi, E. Polturak, and G. Koren, Phys. Rev. Lett. {\bf 84}, 4966 (2000).
\bibitem{Laguna11}  A. Maniv, E. Polturak, and G. Koren, Phys. Rev. Lett. {\bf 91}, 197001 (2003).
\bibitem{Laguna12}  J. Dziarmaga, Phys. Rev. Lett. {\bf 81}, 5485 (1998).
\bibitem{Laguna13}  G. J. Stephens, L. M. A. Bettencourt, and W. H. Zurek, Phys. Rev. Lett. {\bf 88}, 137004 (2002).
\bibitem{Laguna14}  D. Golubchik, E. Polturak, and G. Koren, Phys. Rev. Lett. {\bf 104}, 247002 (2010).
\bibitem{Laguna15}  N. Navon, A. L. Gaunt, R. P. Smith, Z. Hadzibabic, Science {\bf 347}, 167 (2015).
\bibitem{Vinas1} S. Casado, W. Gonz\'{a}lez-Vi\~{n}as, S. Boccaletti, P. L. Ramazza, and H. Mancini, Eur. Phys. J.: Spec. Top. {\bf 146}, 87 (2007).
\bibitem{Vinas2} S. Casado, W. Gonz\'{a}lez-Vi\~{n}as, and H. Mancini, Phys. Rev. E {\bf 74}, 047101 (2006).
\bibitem{Vinas3} M. A. Miranda, D. Laroze, and W. Gonz\'{a}lez-Vi\~{n}as, J. Phys.: Condens. Matter {\bf 25}, 404208 (2013).
\bibitem{Vinas4} M. A. Miranda, J. Burguete, H. Mancini, and W. Gonz\'{a}lez-Vi\~{n}as, Phys. Rev. E
{\bf 87}, 032902 (2013).
\bibitem{Vinas5} M. A. Miranda, J. Burguete, W. Gonz\'{a}lez-Vi\~{n}as, and H. Mancini, Int. J. Bifurcation Chaos Appl. Sci. Eng. {\bf 22}, 1250165 (2012).
\bibitem{Vinas6} W. Gonz\'{a}lez-Vi\~{n}as, S. Casado, J. Burguete, H. Mancini, and
S. Boccaletti, Int. J. Bifurcation Chaos Appl. Sci. Eng. {\bf 11}, 2887 (2001).
\bibitem{Vinas7} D. Ibaceta and E. Calzetta, Phys. Rev. E {\bf 60}, 2999 (1999).
\bibitem{Liu}C.-W. Liu, A. Polkovnikov, and A. W. Sandvik, Phys. Rev. B {\bf 89}, 054307 (2014).
\bibitem{qkz1} W. H. Zurek, U. Dorner, and P. Zoller, Phys. Rev. Lett. {\bf 95}, 105701 (2005).
\bibitem{qkz2} J. Dziarmaga, Phys. Rev. Lett. {\bf 95}, 245701 (2005).
\bibitem{qkz3} A. Polkovnikov, Phys. Rev. B {\bf 72}, 161201(R) (2005).
\bibitem{qkz4} B. Damski and W. H. Zurek, Phys. Rev. Lett. {\bf 99}, 130402 (2007).
\bibitem{qkz5} D. Sen, K. Sengupta, and S. Mondal, Phys. Rev. Lett. {\bf 101}, 016806 (2008).
\bibitem{qkz6} S. Deng, G. Ortiz, and L. Viola, Europhys. Lett. {\bf 84}, 67008 (2008).
\bibitem{qkz7} C. De Grandi, A. Polkovnikov, and A. W. Sandvik, Phys. Rev. B {\bf 84}, 224303 (2011).
\bibitem{qkz8} M. Kolodrubetz, B. K. Clark, and D. A. Huse, Phys. Rev. Lett. {\bf 109}, 015701 (2012).
\bibitem{qkz9} M. Kolodrubetz, D. Pekker, B. K. Clark, and K. Sengupta, Phys. Rev. B {\bf 85}, 100505(R) (2012).
\bibitem{qkz10} S. Yin, X. Qin, C. Lee, and F. Zhong, arXiv:1207.1602.
\bibitem{revqkz1} J. Dziarmaga, Adv. Phys. {\bf 59}, 1063 (2010).
\bibitem{revqkz2} A. Polkovnikov, K. Sengupta, A. Silva, and M. Vengalattore, Rev. Mod. Phys. {\bf 83},
863 (2011).

\bibitem{Cardy} J. Cardy, {\it Scaling and Renormalization in Statistical Physics}, (Cambridge University Press, Cambridge, 1996).
\bibitem{Zhongjp6} F. Zhong, Phys. Rev. E {\bf 73}, 047102 (2006).


\bibitem{Haller} E. Haller, R. Hart, M. J. Mark, J. G. Danzl, L. Reichs\"{o}llner, M. Gustavsson, M. Dalmonte, G. Pupillo, and H.-C. N\"{a}gerl, Nature, {\bf 466}, 597 (2010).
\bibitem{DeGrandi}C. De Grandi, R. A. Barankov, and A. Polkovnikov, Phys. Rev. Lett. {\bf 101}, 230402 (2008).
\bibitem{Dengss} S. Deng, G. Ortiz, and L. Viola, Phys. Rev. B {\bf 83}, 094304 (2011).
\bibitem{Zurek3} A. Das, J. Sabbatini, and W. H. Zurek, Scientific Reports {\bf 2}, 352 (2012).
\bibitem{Zurek5} K. Pyka, J. Keller, H. L. Partner, R. Nigmatullin, T. Burgermeister, D. M. Meier, K. Kuhlmann, A. Retzker, M. B.
Plenio, W. H. Zurek, A. del Campo, and T. E. Mehlst\"{a}ubler, Nat. Commun. {\bf 4}, 2291 (2013).
\bibitem{Polkov1} C. De Grandi, V. Gritsev, and A. Polkovnikov, Phys. Rev. B {\bf 81}, 012303 (2010).
\bibitem{Polkov2} C. De Grandi, V. Gritsev, and A. Polkovnikov,  Phys. Rev. B {\bf 81}, 224301 (2010).
\bibitem{Hohenberg} P. C. Hohenberg and B. I. Halperin, Rev. Mod. Phys. {\bf 49}, 435 (1977).
\bibitem{Tauber} U. C. T\"{a}uber, {\it Critical dynamics: A Field Theory Approach to Equilibrium and Non-equilibrium Scaling Behavior}, (Cambridge University Press, Cambridge, 2014).
\bibitem{Calabrese} P. Calabrese and A. Gambassi, J. Phys. A: Math. Gen. {\bf 38}, R133 (2005).
\bibitem{Janssen1} H. K. Janssen, B. Schaub, and B. Schmittmann, Z. Phys. B {\bf 73}, 539 (1989).
\bibitem{Janssen2} H. K. Janssen, in {\it From Phase Transition to Chaos}, edited by G. Gy\"{o}rgyi, I. Kondor, L. Sasv\'{a}ri, and T. T\'{e}l (World Scientific, Singapore, 1992).

\bibitem{Yin} S. Yin, P. Mai, and F. Zhong, Phys. Rev. B, {\bf 89}, 144115 (2014).
\bibitem{Huse89}D. A. Huse, Phys. Rev. B {\bf 40}, 304 (1989).
\bibitem{Zhong1} S. Gong, F. Zhong, X. Huang, and S. Fan, New J. Phys. {\bf 12}, 043036 (2010).
\bibitem{Zhong2}  F. Zhong, in {\it Applications of Monte Carlo Method in Science
and Engineering}, Edited by S. Mordechai (InTech, Rijeka, 2011).

\bibitem{Yin3} S. Yin, P. Mai, and F. Zhong, Phys. Rev. B, {\bf 89}, 094108 (2014).
\bibitem{Huang} Y. Huang, S. Yin, B. Feng, and F. Zhong, Phys. Rev. B {\bf 90}, 134108 (2014).
\bibitem{Yin2} Q. Hu, S. Yin, and F. Zhong, Phys. Rev. B, {\bf 91}, 184109 (2015).
\bibitem{Chandran} A. Chandran, A. Erez, S. S. Gubser, and S. L. Sondhi, Phys. Rev. B {\bf 86}, 064304 (2012).
\bibitem{Zheng} B. Zheng, Phys. Rev. Lett. {\bf 77}, 679 (1996).
\bibitem{Yin4}S. Zhang, S. Yin, and F. Zhong, Phys. Rev. E {\bf 90}, 042104 (2014).
\bibitem{Ito} Y. Ozeki and N. Ito, J. Phys. A {\bf 40}, R149 (2007).

\bibitem{Li1} Z. Li, U. Ritschel, and B. Zheng, J. Phys. A {\bf 27}, L837 (1994).
\bibitem{Li2} E. V. Albano, M. A. Bab, G. Baglietto, R. A. Borzi, T. S. Grigera, E. S. Loscar, D. E. Rodriguez, M. L. R. Puzzo, and G. P. Saracco, Rep. Prog. Phys. {\bf 74}, 026501 (2011).
\bibitem{ZhengB} B. Zheng, Int. J. Mod. Phys. B {\bf 12}, 1419 (1998).
\bibitem{Binder} D. P. Landau and K. Binder, {\it A Guide to Monte Carlo Simulations in Statistical Physics}, 3rd edition (Cambridge University Press, Cambridge, 2009).
\bibitem{Farhi} E. Farhi, J. Goldstone, S. Gutmann, and M. Sipser, arXiv:quant-ph/0001106v1 (2000).
\end{thebibliography}
\end{document}